\documentclass{vgtc}                          

\ifpdf
  \pdfoutput=1\relax                   
  \usepackage{graphicx}                
  \DeclareGraphicsExtensions{.pdf,.png,.jpg,.jpeg} 
\else
  \usepackage{graphicx}                
  \DeclareGraphicsExtensions{.eps}     
\fi%

\graphicspath{{figures/}{pictures/}{images/}{./}} 

\usepackage{microtype}                 
\PassOptionsToPackage{warn}{textcomp}  
\usepackage{textcomp}                  
\usepackage{mathptmx}                  
\usepackage{times}                     
\usepackage{cite}                      
\usepackage{tabu}                      
\usepackage{booktabs}                  
\usepackage{balance}

\onlineid{0}

\vgtccategory{Research}

\vgtcinsertpkg

\title{A Visual Analytics Approach to Scheduling Customized Shuttle Buses via Perceiving Passengers' Travel Demands}

\author{Qiangqiang Liu\thanks{$\lbrace$alfredliu,forrestli,jsontang,bindylin,tobychen$\rbrace$@webank.com} %
\and Quan Li$^{*}$\thanks{corresponding author} %
\and Chunfeng Tang$^{*}$
\and Huanbin Lin$^{*}$
\and Xiaojuan Ma\thanks{mxj@cse.ust.hk}
\and Tianjian Chen$^{*}$}
\affiliation{\scriptsize $^{*}$AI Group, WeBank, Shenzhen, Guangdong, China $\quad$ $^{\dag}$The Hong Kong University of Science and Technology, Hong Kong}

\teaser{
  \centering
      \vspace{-5mm}
  \includegraphics[width=\linewidth]{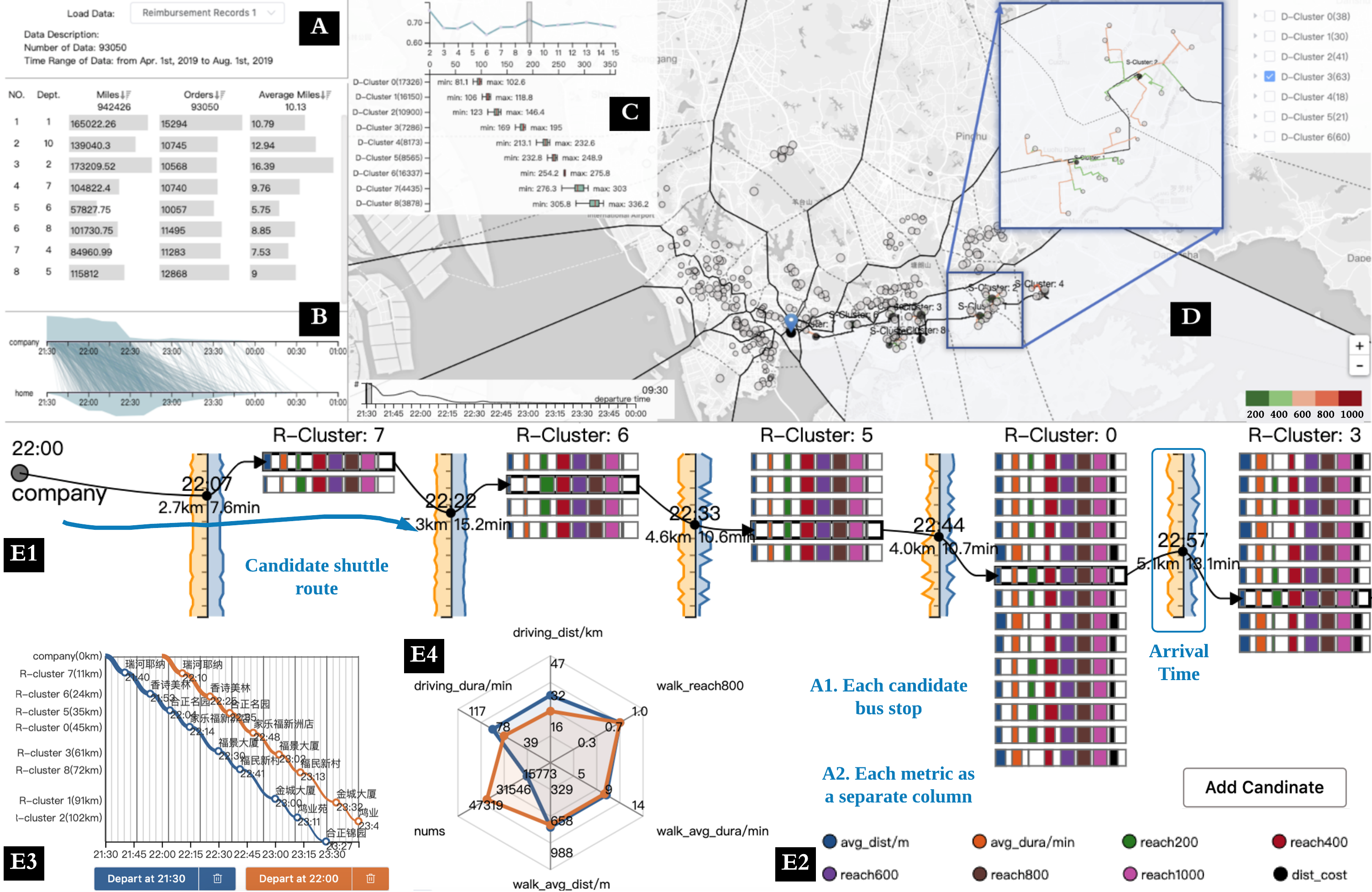}
      \vspace{-7mm}
   \caption{\textit{ShuttleVis} includes (A) a dataset loader and data description; (B) overview of car-hailing reimbursement records across different departments and descriptions of the departure and arrival time; (C) directional clustering configuration view to help analysts identify appropriate travel directions; (D) map view to visualize identified directional and regional clustering results, and comparative ranking view that includes (E1) a ranking of shuttle bus stops in terms of (E2) metrics in consecutive regional clusters along one travel direction, (E3) timetables of selected shuttle routes, and (E4) radar chart showing attribute distributions of selected routes.}
  \label{fig:teaser}
        \vspace{-2mm}
}

\abstract{Shuttle buses have been a popular means to move commuters sharing similar origins and destinations during periods of high travel demand. However, planning and deploying reasonable, customized service bus systems becomes challenging when the commute demand is rather dynamic. It is difficult, if not impossible to form a reliable, unbiased estimation of user needs in such a case using traditional modeling methods. We propose a visual analytics approach to facilitating assessment of actual, varying travel demands and planning of night customized shuttle systems. A preliminary case study verifies the efficacy of our approach.
} 

\CCScatlist{
  \CCScatTwelve{Human-centered computing}{Visu\-al\-iza\-tion}{Visualization design and evaluation methods}{}
}

\begin{document}


\firstsection{Introduction}

\maketitle

\par As a kind of ``Mobility as a Service'' (MaaS)~\cite{liu2015analysis}, shuttle buses that travel along a fixed route following a pre-determined commuting schedule have been a popular means to move commuters sharing similar origins and destinations during periods of high travel demand, due to its advantages of congestion alleviation, environment friendliness, and better user experience~\cite{lyu2019cb}. However, planning shuttle bus routes in a reliable and cost-efficient manner is nontrivial due to the following three challenges: \textit{(1) Unreliable and biased collection of travel demands.} Shuttle bus service providers mainly rely on online surveys to gather travel demands of potential passengers and further aggregate similar demands to generate candidate bus routes. This strategy is inflexible since survey results only represent the views of respondents -- a limited sample of potential commuters, leading to possible biases in the data that may not well reflect the reality~\cite{lefever2007online}. As a result, current shuttle service cannot sufficiently fulfill the needs of many commuters, either because of misaligned schedule or inconvenient drop-off spots. \textit{(2) Labor-intensive planning and design.} It is very tedious, inefficient, and costly to plan customized commuter shuttle routes and schedules by manually analyzing the collected travel demands from online survey results~\cite{liu2015analysis}. It takes a long time from the initial data collection to the final deployment of a new route. \textit{(3) Dynamic and varying travel demands.} It is hard to detect changes in travel demands and patterns and adjust the shuttle service accordingly and timely by only leveraging and manually processing the results of the online surveys. Furthermore, maintaining a reasonable route and schedule requires a comprehensive consideration of a variety of factors that may affect the transit experience. Otherwise, shuttle service providers would have to make judgment calls eventually. In this study, we collaborate with domain experts and identify their primary needs and concerns by first exploiting travel reimbursement data to derive commuters' needs overlooked by the current shuttle services. Then we enable shuttle service providers to explore possible travel direction and destination clustering for determining appropriate shuttle bus bounds for different areas and stops which optimize the reachability to commuters' home. Based on the results, we generate and compare candidate bus routes and schedules by considering the factors that may affect the transit experience such as the departure time distribution, drop-off spot distribution, walking distances to destinations, and traffic conditions. An in-depth case study is conducted to evaluate the efficacy of our approach.

\section{Related Work}

\par Understanding \textbf{Human Mobility Pattern} has significant impacts and enables various applications such as urban planning~\cite{zheng2011urban}, region function analysis~\cite{yuan2012discovering,wu2020towards}, hotspots detection~\cite{zheng2009mining}, driving route recommendation~\cite{yuan2011t,ge2010energy}, and bus route planning~\cite{bastani2011greener,chen2013b,2018Solving,2020Graph}. Bastnai et al.~\cite{bastani2011greener} grouped taxi trips into clusters and identified a route connecting multiple clusters to maximize the sum of each connected trip cluster without considering other constraints such as time. Chen et al.~\cite{chen2013b} clustered ``hot'' areas with dense pick-up/drop-off to identify a candidate bus stop and then derived several rules to automatically generate candidate bus routes. The above work solve issues that contain multiple origins and destinations and make certain assumptions in problem formulation and evaluation such as fixed walkable distance and threshold for cluster split, which may affect the results and subsequent decisions. We focus on one origin to multiple destinations and allow shuttle service providers to explore travel directions and destination clusters that optimize walking reachability to destinations. Kamw et al.~\cite{2019Urban} proposed a computational model to help users study the jointly constrained accessible regions, street segments, and Points of Interest (POIs), while our work focuses on the reachability between selected shuttle stop and home destinations.

\par \textbf{Vehicle Routing Problem} is intensively studied in urban planning and transportation field~\cite{chen2013b}. Bus network design determines routes and operation frequencies to achieve certain objectives, e.g., the shortest route, shortest travel time, subject to certain constraints. However, objectives selection should consider both operators' and passengers' requirements which are often conflicting, leading to a design trade-off rather than an optimal solution~\cite{weng2020pareto}. Early bus network design is mainly based on surveys to get travel demands and passenger flows~\cite{chua1984planning,guihaire2008transit}. Some work~\cite{chen2013b,lyu2019cb} leverage passenger OD flows to find a bus route with a fixed frequency, maximizing the number of passengers along a fixed route subject to the total travel time constraint. Different from purely automatic methods, we allow analysts to interactively configure travel directions and regional clusters based on walking reachability. We also compare different routes and consider factors that may affect transit experience.

\par \textbf{Traffic Data Visualization} mainly deals with three types of traffic data: event-, location-, and movement-based data~\cite{wang2014visual}, and various applications have been developed~\cite{chen2015survey,ferreira2013visual,liu2013vait,wang2014visual,chu2014visualizing,guo2011tripvista,zeng2013visualizing,andrienko2009interactive,rinzivillo2008visually,wang2013visual,pu2013t,piringer2012alvis,anwar2014traffic}. For example, Schreck et al.~\cite{schreck2009visual} combined automatic data analysis and human supervision. Users can monitor and control the clustering process and attain appropriate results. Users can also interactively refine clustering results such as excluding sub-clusters from a cluster or dividing a cluster into several smaller sub-clusters~\cite{andrienko2009interactive,rinzivillo2008visually}. Similarly, we incorporate human intelligence in the analysis loop to interactively initialize directional and regional clusters.

\section{Observational Study and Requirements}
\par We worked with a team of domain experts from the Department of Accounting and Human Resources at WeBank (an internet bank), including a chief director (E.1, male, age: $37$), a financial director (E.2, male, age: $31$), a human resource personnel (E.3, female, age: $26$) and a shuttle bus operation manager (E.4, male, age: $35$) to identify their primary concerns about making plans of customized shuttle bus routes. Typically, they planned customized commuter buses in four stages. First, passengers who are willing to take the bus needed to submit a travel demand survey. Second, E.1 and E.3 collected the travel demand data and classified the passengers with similar travel destinations to obtain a candidate customized bus route according to OD traveling directions. Third, E.4 made a field survey for each alternative shuttle route to check road conditions and adjusted the initial design of the shuttle route accordingly. Finally, the actual trial operation of the customized shuttle bus route was carried out and the running time of the route was estimated for scheduling. At present, the number of the customized commuter bus routes during the daytime is $18$. However, with respect to the daytime shuttle buses, the night shuttle buses for the overtime has not been well implemented. The majority of the employees who work overtime (i.e., after 21:30) prefer to take a taxi on their own and the number of employees who voluntarily submit overtime travel demand questionnaires is very limited, leading to a large number of unmet potential travel demands. E.2 commented that if they directly follow the regular shuttle bus routes during the daytime for night operation, the routes may be biased due to different demands and traffic conditions. In other words, the conventional perception and processing of the travel demands would inevitably affect the bus schedules and routes. To sum up, we need to meet the following requirements: \textbf{R.1 Understand Employees' Travel Demands for Overtime.} Offering night customized bus routes needs to have a clear overview of employees' travel demands for overtime. \textbf{R.2 Generate and Compare Shuttle Bus Routes.} The experts required an interactive visual exploration that considers the factors that may affect the transit experience within a predictable time duration combining with automatic recommendation of bus routes that connect the origin and a sequence of bus stops to the destination. \textbf{R.3 Compare Candidate Shuttle Bus Stops.} How to optimize the reachability to commuters' home destinations and deliver the best walking experience is a great concern for placing the candidate customized shuttle bus stops.

\section{Data Processing and System Overview}
\par The internal shuttle bus system for the daytime and reimbursement financial system launched at WeBank provided us with several datasets, which record three types of information: \textbf{(1) Urban Road Network Data.} The local urban road network data comprises a directed graph of the city of \textit{Shenzhen}, of which the vertices represent road intersections and the edges represent roads. Specifically, this graph has $253,890$ vertices and $314,234$ edges; \textbf{(2) Car-hailing Reimbursement Data.} We collected $93,050$ employees' car-hailing reimbursement records for overtime with de-identification, which ranges from April $1^{st}$ to August $1^{st}$, 2019. Each car-hailing reimbursement record includes the \textit{employee ID}, \textit{departure time}, \textit{arrival time}, \textit{place of origin}, \textit{place of destination}, and \textit{payment amount}. Particularly, the place of origin is the location of the company and the places of destination are the employees' residential locations. We first unified similar residential areas into the same locations since they may differ in the building No.(s) or entrances of residential districts. We then obtained their geographical representations (i.e., latitude-longitude) for all the locations through an online map API, followed by a manual calibration; \textbf{(3) Daytime Shuttle Routes and Timetables.} As mentioned, there are $18$ shuttle routes in the daytime picking up passengers with fixed timetables. Furthermore, to obtain the traveling time and route from one bus stop to another starting at different departure timestamps, we obtain a dataset \textbf{(D$_{21:30}$)} by dynamically crawling traffic conditions and the recommended routes between two stops. To avoid data biases, we obtained the data of every weekday and conducted necessary calibration of traveling routes and average operation of traveling duration. Specifically, starting from 21:30 with an interval of 5 minutes, we obtained the recommended bus routes from the workplace to the first bus stop and the routes between the first and second stop with an interval of 1 minute, and so on. Following the criteria of setting up shuttle routes, i.e., \textit{move forward}, \textit{destination-closer}, \textit{no zigzag routes}~\cite{chen2013b}, we developed \textit{ShuttleVis} (\autoref{fig:teaser}) and conducted the following steps to allow shuttle service providers to generate and adjust shuttle bus stops, routes, and schedules.

\par
\begin{figure}[h]
\centering
\vspace{-3mm}
  \includegraphics[width=\linewidth]{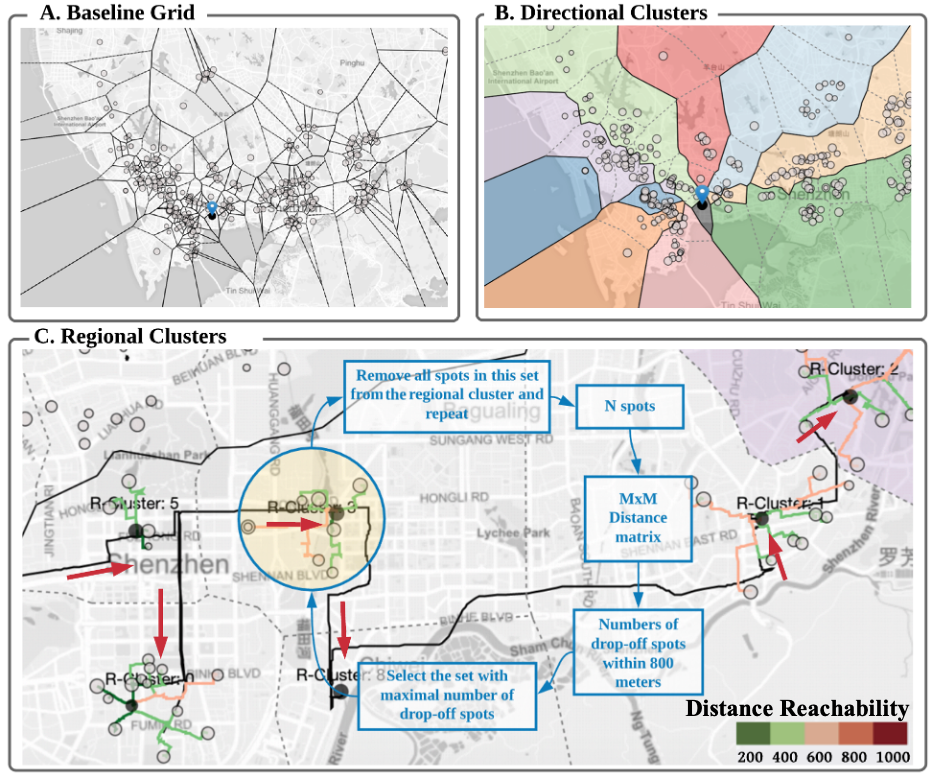}
    \vspace{-6mm}
  \caption{(A) Voronoi grid-based map design on the basis of drop-off spots. (B) Directional clusters highlighted by different colors and regional clusters separated by dashed lines. (C) The procedure for generating regional clusters. We string (indicated by red arrows) and link consecutive shuttle stops to each destination.}
  \label{fig:map}
  \vspace{-2mm}
\end{figure}

 \textbf{Step \raisebox{.5pt}{\textcircled{\raisebox{-.9pt} {1}}} Determine Travel Directions via Clustering Configuration View (R.1).} The first step is to determine the travel directions through directional clustering. We use K-means to obtain the initial number of travel directions. We determine the value of K by finding the peak point in the relationship between the cluster number and the \textit{Silhouette Coefficient}~\cite{ROUSSEEUW1987Silhouettes} among the generated clusters. As shown in \autoref{fig:teaser}(C), the \textit{x}-axis represents the number of travel directions and the \textit{y}-axis indicates the value of the \textit{Silhouette Coefficient} corresponding to the number of travel directions. It can be witnessed that the \textit{Silhouette Coefficient} attains the highest value when the number of travel directions is $2$. However, determining the number of travel directions faces more practical issues. The distribution range of the angle towards the company of the home destinations may expand too much to set up only one route that connects all the destinations. To handle this issue, we visualize the angle distribution of the corresponding home destination towards the company as a box plot for each travel direction, from which the experts could clearly observe the angle distribution in the current condition. In other words, the more concentrated the box plot, the more concentrated the angles of the destinations towards the company. Taken together, we choose $9$ as the directional clustering number in this case which has a reasonable angle distribution and distribution of reimbursement records in each directional cluster. Based on the above processing, each car-hailing record has a unique directional cluster id. The result of directional clustering is shown as the colored parts in \autoref{fig:map}(B).

\par \textbf{Step \raisebox{.5pt}{\textcircled{\raisebox{-.9pt} {2}}} Initialize Shuttle Stops by Regional Clustering for Each Travel Direction via Map View (R.1).} For each travel direction, we use regional clustering to initialize the candidate shuttle bus stop that can cover nearby drop-off spots within one cluster for each travel direction. \autoref{fig:map}(C) shows how we generate the regional clusters for each directional cluster. We first iterate all drop-off spots $N$. For each drop-off spot $n_i$, we construct a walking distance matrix with the dimension of $M \times M$. We predefine the walking distance between the two spots in each pair of the drop-off spot destinations in this matrix is within $1000$ meters after discussing with the experts. In other words, $M$ is less than $N$ since the distance between any two spots can be larger than $1000$ meters. Therefore, each drop-off spot destination corresponds to a set in which the walking distance between any two spots is within $1000$ meters. We choose the set with the maximal number of drop-off spots as the first regional cluster. Next, we remove all the corresponding drop-off spots within this regional cluster and repeat the above procedures. Thus, each car-hailing reimbursement record then has a unique directional and regional cluster id. We propose a Voronoi grid-based map view to help experts understand the identified directional and regional clustering results. We first construct Voronoi grids~\cite{aurenhammer2000voronoi} as the base for our map design on the basis of the drop-off spots extracted from the car-hailing reimbursement records (\autoref{fig:map}(A)). Two reasons are considered for choosing the Voronoi grid~\cite{Siming2017E}. First, the polygon in Voronoi is irregular, similar to the real-world terrain. Second, the neighborhood representation is convenient to maintain based on Delaunay Triangulation. After constructing Voronoi grids, we use different visual cues to represent the boundaries of regional clusters that belong to the same or different directional clustering. Particularly, for a specific Voronoi edge, if the shuttle bus stop on each side of this edge belongs to different directional clusters, we depict this edge as a solid line; otherwise, if the shuttle bus stop on either side of this edge belongs to a same directional cluster but different regional clusters, we depict this edge as a dashed line. In other cases, we just remove the edge (\autoref{fig:map}(B)).

\begin{figure*}[h]
\centering
\vspace{-6mm}
  \includegraphics[width=\linewidth]{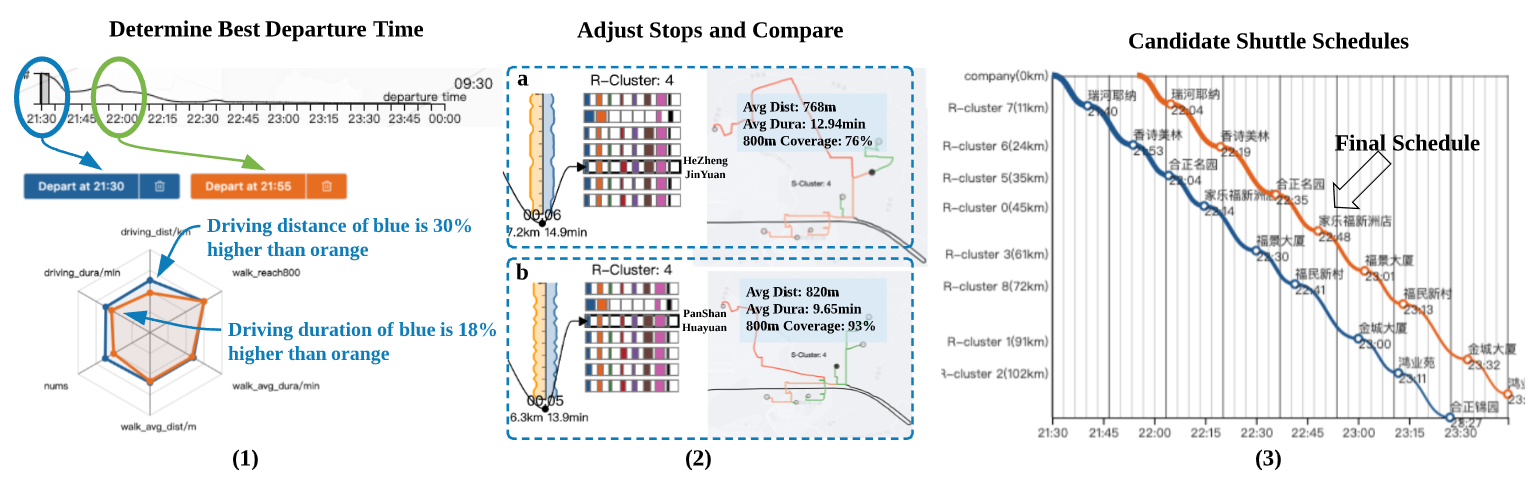}
    \vspace{-8mm}
  \caption{Experts selected two departure timestamps 21:30 and 21:55 and compared the two routes in terms of (1) different metrics. (2) They adjusted the shuttle bus stop from ``\textit{HeZhengJinYuan}'' to ``\textit{PanShanHuaYuan}''  and (3) finalized the route that starts at the departure time of 21:55.}
  \label{fig:case1}
      \vspace{-6mm}
\end{figure*}

\par \textbf{Step \raisebox{.5pt}{\textcircled{\raisebox{-.9pt} {3}}} Recommend Shuttle Routes and Walking Paths (R.2).} Up to this point, we obtain a list of candidate shuttle bus stops along one travel direction via directional and regional clustering. The next step is to recommend a shuttle route that strings those consecutive shuttle stops according to the distance to the workplace. To intuitively represent the shuttle route for each directional cluster and the path from the initialized shuttle stop to its covered home destinations, we estimate the shuttle routes and walking paths to each home destination on the basis of the previously mentioned dataset \textbf{(D$_{21:30}$)}. As shown in \autoref{fig:map}(C), the black line shows a complete shuttle route and multiple lines between the shuttle stop (black points) and home destinations (grey points) indicate the walking paths with color indicating different distance reachability.

\par \textbf{Step \raisebox{.5pt}{\textcircled{\raisebox{-.9pt} {4}}} Refine Shuttle Bus Stops and Routes via Comparative Ranking View (R.3).} Although the map view conveys information about the candidate shuttle bus stops and routes, the domain experts need to explore the properties of each candidate shuttle bus stop to assess its reachability. After discussing with the experts, we consider the following factors that may affect the transit experience to evaluate a candidate shuttle bus stop (\autoref{fig:teaser}(E2)). \textit{(1) avg\_dist:} the weighted average distance from one selected shuttle bus stop to the other destinations in the same regional cluster; \textit{(2) avg\_dura:} the weighted average walking duration from one selected shuttle bus stop to the other destinations in the same regional cluster. Note that \textit{weighted} means that we consider the passenger number at each drop-off spot when calculating the \textit{avg\_dist} and \textit{avg\_dura}. \textit{(3) reach200:} the ratio of the number of car-hailing orders from one shuttle bus stop to the other destinations within 200m to all the car-hailing orders in the same regional cluster. The definition also applies to \textit{reach400}, \textit{reach600}, and so on; \textit{(4) dist\_cost:} we obtain $cost_i$ by multiplying the distance from one selected shuttle bus stop $i$ to the other destinations in the same regional cluster by the number of the orders and \textit{dist\_cost} is just the accumulative value of $cost_i$. 

\par Inspired by \textit{EmbeddingVis}~\cite{li2018embeddingvis} and Marey's train schedule~\cite{marey1878methode}, as shown in \autoref{fig:teaser}(E1), we present each candidate shuttle bus stop as a combined bar, in which the length of a single bar with colors indicates the normalized metric value of the corresponding shuttle bus stop. We line up all the regional clusters in one travel direction horizontally and connect selected shuttle bus stops across all the regional clusters with a curve which forms one candidate shuttle route. The ranking in each regional cluster is based on the value of a certain metric. The left and right vertical curves between two regional clusters indicate the distance between two R-Clusters and arrival time starting at different departure timestamps by using the dataset of \textbf{(D$_{21:30}$)}, respectively. Experts can select multiple routes (up to three) for simultaneous comparison by adding them to a candidate list and they would appear in the radar chart (\autoref{fig:teaser}(E4)). \textit{Driving\_dura}, \textit{driving\_dist}, \textit{walk\_reach800}, \textit{walk\_avg\_dura}, \textit{walk\_avg\_dist}, and \textit{nums} indicate the driving duration, distance, the ratio of walking reachability within 800m, the average duration and walking distance, and the number of involved car-hailing records around the corresponding departure timestamp of the selected route, respectively. A timetable of the selected routes is shown in \autoref{fig:teaser}(E3), with \textit{x}-axis representing the arrival time and \textit{y}-axis representing the regional clusters and distance to the workplace. Experts can choose different departure timestamps and \textit{ShuttleVis} automatically recommends a new route.

\section{Case Study}
\par
\begin{figure}[h]
\centering
      \vspace{-2mm}
  \includegraphics[width=\linewidth]{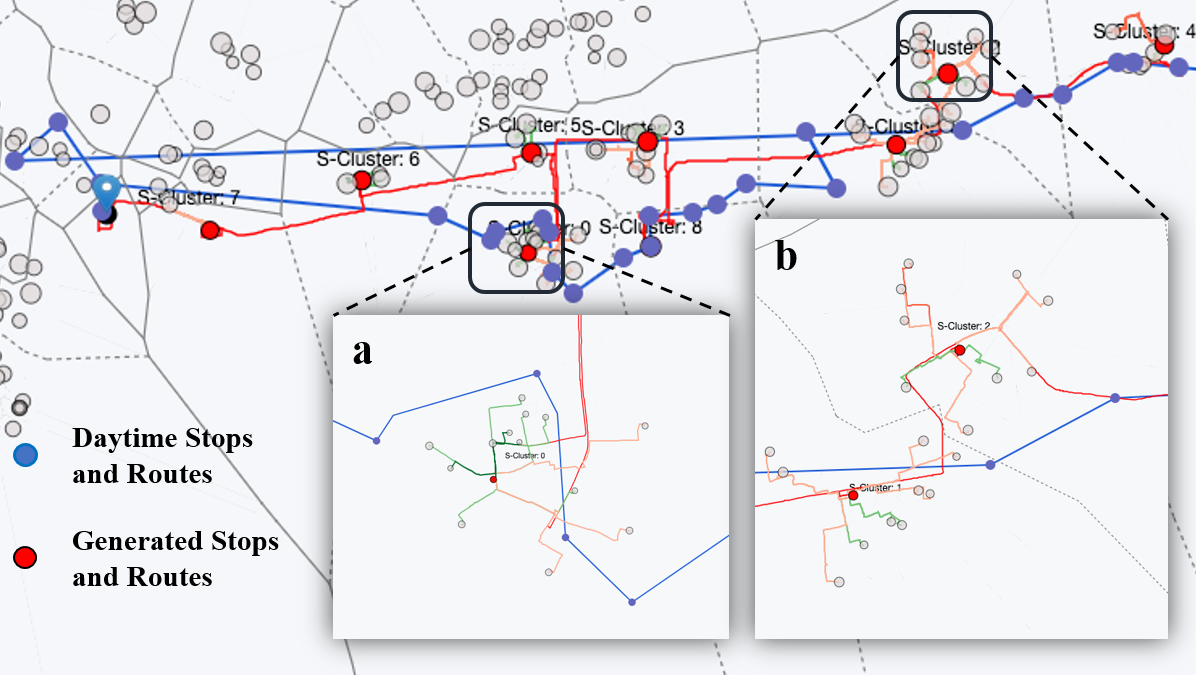}
    \vspace{-6mm}
  \caption{The route adjusted by \textit{ShuttleVis} and that in the daytime.}
  \label{fig:case2}
      \vspace{-4mm}
\end{figure}

 We introduce a case conducted by E.1 and E.4, who used \textit{ShuttleVis} to study the night customized shuttle bus operation. After loading the dataset and \textbf{setting the number of directional clusters} to $9$ as previously mentioned, they chose directional cluster $3$ for further inspection. First, they needed to \textbf{select an appropriate departure time} for the shuttle which could cover as many passengers as possible. They witnessed that there are two peaks corresponding to the departure time of 21:30 and 21:55 (\autoref{fig:case1}(1)), respectively, so they added them to the candidate list for exploration. From the radar chart in \autoref{fig:case1}(1), they observed that although the route starting at 21:30 covers more passengers (indicated by the \textit{nums} axis) than the other route, the driving distance and duration of the route starting at 21:30 are higher than that of the route starting at 21:55, i.e., the driving distance and the driving duration of the blue route is 30\% and 18\% higher than that of the orange route, respectively. E.1 commented that their company provides extra transportation compensation for their employees who work after 21:30 and thought this finding makes sense since there should be more passengers calling for car-hailing at 21:30, which may lead to traffic congestion. The experts then suggested that the departure time for the night shuttle bus should be 21:55. After determining the departure timestamp, the experts moved to \textbf{verify the recommended shuttle stops and routes}. Note that \textit{ShuttleVis} recommends the default shuttle stops and routes based on \textit{avg\_dist}. They observed that in R-Cluster 4, the system recommends ``\textit{HeZhengJinYuan}'' as the shuttle bus stop but the experts identified that another drop-off spot ``\textit{PanShanHuaYuan}'' is located in the middle part in this regional cluster (\autoref{fig:case1}(2)). To obtain more information, E.1 clicked on the rectangle of ``\textit{PanShanHuaYuan}'' in \autoref{fig:case1}(2)(b) and found that the \textit{avg\_dist}, \textit{avg\_dura}, and \textit{coverage within 800m} is 820m, 9.65min, and 93\%, respectively, compared with that of ``\textit{HeZhengJinYuan}'' of which the \textit{avg\_dist} is 768m but with a longer \textit{avg\_dura} (12.94min) and smaller coverage (76\%). E.4 reported that this may be due to footbridge or subway (i.e., shorter distance but longer duration). Therefore, they decided to change the shuttle stop to ``\textit{PanShanHuaYuan}'' for this regional cluster. To further inspect differences between daytime and night, they \textbf{compared the routes by \textit{ShuttleVis} for overtime with the daytime route}. As shown in \autoref{fig:case2}, the red line indicates the stops and route by \textit{ShuttleVis} while the blue one indicates the daytime stops and route. Generally, the blue route connects every drop-off spot but cannot well cover drop-off spots at night (\autoref{fig:case2}b). In \autoref{fig:case2}a, although the daytime blue route covers most night drop-off spots, it potentially increases the driving duration.

\section{Conclusion and Future Work}
\par In this study, we introduce a visual analytics approach \textit{ShuttleVis} to facilitating assessment of actual, varying travel demands and planning of customized night shuttle buses. It allows shuttle service providers to explore traveling directional and regional clustering that optimize the reachability to commuters' home destinations. Based on the identified directions and bus stops, candidate shuttle routes and schedules are provided for comparison by considering the factors that may affect the transit experience. In the future, we shall continue perceiving travel demands by considering other passenger flows and it may be necessary to set up multiple shuttle routes for one direction in different periods of time if there are many requests.

\acknowledgments{
We thank the anonymous reviewers for their valuable comments. This research was supported in part by HKUST - WeBank Joint Laboratory Project Grant No.: WEB19EG01-d.}

\balance
\bibliographystyle{abbrv-doi}

\bibliography{template}
\end{document}